\documentclass[12pt]{article}

\def\square{\kern1pt\vbox{\hrule height 1.2pt\hbox{\vrule width 1.2pt\hskip 3pt
    \vbox{\vskip 6pt}\hskip 3pt\vrule width 0.6pt}\hrule height 0.6pt}\kern1pt}
\def\gtwid{\mathrel{\raise.3ex\hbox{$>$\kern-.75em\lower1ex\hbox{$\sim$}}}}
\def\ltwid{\mathrel{\raise.3ex\hbox{$<$\kern-.75em\lower1ex\hbox{$\sim$}}}}

\begin{document}

\begin{titlepage}

\begin{flushright}
CRETE-07-11 \\ UFIFT-QG-07-02
\end{flushright}

\begin{center}
{\bf Reply to `Can infrared gravitons screen $\Lambda$?'}
\end{center}

\begin{center}
N. C. Tsamis$^{\dagger}$
\end{center}

\begin{center}
\it{Department of Physics, University of Crete \\
GR-710 03 Heraklion, HELLAS}
\end{center}

\begin{center}
R. P. Woodard$^{\ddagger}$
\end{center}

\begin{center}
\it{Department of Physics, University of Florida \\
Gainesville, FL 32611, UNITED STATES}
\end{center}

\vspace{1cm}

\begin{center}
ABSTRACT
\end{center}
We reply to the recent criticism by Garriga and Tanaka of our proposal
that quantum gravitational loop corrections may lead to a secular
screening of the effective cosmological constant. Their argument rests
upon a renormalization scheme in which the composite operator $(R \sqrt{-g}
- 4 \Lambda \sqrt{-g}\, )_{\rm ren}$ is defined to be the trace of the
renormalized field equations. Although this is a peculiar prescription,
we show that it {\it does not preclude secular screening}. Moreover,
we show that a constant Ricci scalar {\it does not even classically} 
imply a constant expansion rate. Other important points are: (1) the 
quantity $R_{\rm ren}$ of Garriga and Tanaka is neither a properly 
defined composite operator, nor is it constant; (2) gauge dependence 
does not render a Green's function devoid of physical content; (3) 
scalar models on a non-dynamical de Sitter background (for which there 
is no gauge issue) can induce arbitrarily large secular contributions 
to the stress tensor; (4) the same secular corrections appear in 
observable quantities in quantum gravity; and (5) the prospects seem 
good for deriving a simple stochastic formulation of quantum gravity 
in which the leading secular effects can be summed and for which the 
expectation values of even complicated, gauge invariant operators can 
be computed at leading order.

\begin{flushleft}
\end{flushleft}

\begin{flushleft}
$^{\dagger}$ e-mail: tsamis@physics.uoc.gr \\
$^{\ddagger}$ e-mail: woodard@phys.ufl.edu
\end{flushleft}

\end{titlepage}

\section{Introduction}

Some years ago we proposed that the continual production of infrared 
gravitons induces a sort of quantum friction which gradually slows 
inflation \cite{TW1}. When inflation rips a pair of infrared gravitons 
from the vacuum, which is a 1-loop effect, they induce a gravitational 
potential that contributes to the energy density at next order. This
potential remains imprinted on the spacetime even after the pair has been 
pulled out of causal contact. The number density of pairs remains constant
because the vast expansion of the spatial volume cancels the continual 
creation of pairs.  However, the induced gravitational potential grows 
without bound. Taking account of the number of infrared gravitons inside 
the past light-cone of a local observer gives $\Phi \sim -G H^2 \cdot H t$ 
\cite{TW2}. The effect of this potential must be to slow inflation 
because gravity is attractive. Hence the induced energy density of 
interaction is negative, $\rho \sim -G H^6 \cdot Ht$. Because gravity is 
weak, this energy density grows very slowly, in other words, $\vert 
\dot{\rho} \vert \ll H \vert \rho\vert$ for $H t \gg 1$. Combining that 
fact with stress-energy conservation, $\dot{\rho} = - 3H (\rho + p)$, 
implies the induced pressure is $p \sim - \rho$. Hence one gets a growing, 
negative vacuum energy.

It is tempting to speculate that this mechanism might simultaneously explain
why the observed cosmological constant is so much smaller than the natural
scales of particle physics, and also provide a model of inflation which has 
no fundamental scalar. The idea is that primordial inflation would start 
because the bare cosmological constant is positive and of GUT scale. This
avoids the problem of needing a scalar inflaton to be unnaturally homogeneous 
over a super-Hubble volume \cite{Vach}. Inflation would persist for a long 
time because gravity is a weak interaction, thereby dispensing with the need 
for a shallow potential. And inflation would be brought to an end by the 
gradual accumulation of negative vacuum energy.

Explicit computations in this scheme are challenging because the first
effect should be at two loops and because the putative effect would be 
non-perturbatively strong during the current epoch. Ford was early able to 
show that there is no secular deviation from de Sitter background at one 
loop \cite{Ford}, in agreement with the mechanism. This 1-loop result 
has been confirmed by two other groups \cite{Fabio,TW3}. A year-long 
computation of the graviton 1-point function revealed secular slowing at 
the expected 2-loop order \cite{TW4}.\footnote{This work has not been 
confirmed, nor re-done using dimensional regularization \cite{TW3}.}
Because back-reaction should be small until the last few e-foldings, it 
was possible to use the 2-loop result to estimate the scalar and tensor 
power spectra \cite{ATW}. Some thought has also been given to how the 
mechanism might operate after it becomes non-perturbatively strong at the
end of inflation \cite{TW5}.

Our proposal has recently been criticized by Garriga and Tanaka \cite{GarTan}.
They raise two objections:
\begin{enumerate}
\item{The graviton 1-point function cannot show slowing, or anything else,
because it is gauge dependent; and}
\item{There can be no slowing of inflation in pure quantum gravity because 
the field equations set the renormalized Ricci scalar to a constant.}
\end{enumerate}
In section 2 we point out that physical information can reside in gauge 
dependent quantities, that the secular growth we found was not injected 
through our gauge fixing functional, that the only invariant quantities so 
far checked in quantum gravity show the same type of secular dependence, 
that this secular dependence also affects the vacuum energy of scalar 
models for which there is no gauge issue, and that it could only drop out 
of the quantum gravitational vacuum energy through an infinite series of 
cancellations. Section 3 contains two important points: 
\begin{enumerate}
\item{Garriga and Tanaka did not show that the renormalized Ricci scalar 
is constant and the relation they actually derived is completely consistent 
with secular slowing; and}
\item{They did not define the Ricci scalar as a properly renormalized
composite operator but rather the Ricci scalar times the volume element.}
\end{enumerate}
We also comment on some peculiarities of their renormalization scheme. 
Section 4 makes the short but crucial point that a constant Ricci scalar
does not even classically imply expansion. In Section 5 we review recent 
progress in developing a computational formalism which is powerful enough 
to evolve into the non-perturbative regime.

\section{Gauge Dependence}

The fact that the expectation value of the metric cannot be formed into a 
gauge invariant quantity in the same way as a classical metric was originally
pointed out by Unruh \cite{Unruh}. We agree with him, and two years ago we 
advocated that the quantum gravitational back-reaction on $\Lambda$-driven 
inflation should be quantified with an invariant operator which measures 
the geodesic deviation \cite{TW6}. The expectation value of this operator 
has not yet been computed at 2-loop order owing to its complexity. It is 
worth noting that the difficulty of working with gauge invariant observables 
has led Losic and Unruh as well to the consideration of gauge dependent 
quantities in their own interesting study of back-reaction \cite{LU}. \\

$\bullet$ {\it Gauge Dependence Does Not Automatically Imply Lack Of
Physical Content:} A good example is provided by the 1PI functions of the 
Standard Model. These 1PI functions are certainly gauge dependent, yet 
they can be assembled to give the gauge independent S-matrix, which has 
of course been subject to impressive experimental verification. So the 
gauge-dependent 1PI functions of the Standard Model must contain valid, 
physical and gauge-independent information along with some unphysical and 
gauge-dependent behaviour.

An especially noteworthy example is the gauge dependence of effective 
potentials \cite{Jackiw}. In the days when people were studying the 
possibility for loop corrections to stabilize Kaluza-Klein compactifications
\cite{CW} this gauge dependence was regarded with great suspicion and even 
despair. The question was asked, ``which result should be trusted if the 
effective potential shows a non-trivial minimum in one gauge but not in
another?'' It was even believed that a ``field-space metric'' could be 
identified that would allow construction of a gauge independent (and also 
field re-definition independent) effective action \cite{Sergei,Toms}.

It was eventually realized that gauge dependence is actually a blessing
for this problem. What one really wants is to find a stationary point of
the full effective action; it is only because the effective action is too
complicated to evaluate for an arbitrary background field that one is
reduced to studying the effective potential. Because different gauges are 
related by field-dependent gauge transformations, the same finite-parameter 
family of field configurations in one gauge can, in another gauge, probe a 
different direction in the full space of fields. Hence the correct conclusion 
is not that both effective potentials are rubbish on account of their gauge
dependence, or that one is right and the other wrong, but rather that {\it
each one represents valid physical information about the theory}. For a
putative solution to be correct it must make each of the effective 
potentials stationary --- and even then it might still be spurious if
there is a non-stationary direction that neither of the restricted probes 
of field space chances to access.

The expectation value of the graviton 1-point function in a fixed gauge is 
a perfectly valid measure of quantum corrections to the background {\it in 
that gauge}. It might be that the secular screening we found is an artifact
of the gauge (or even an error in what was a very long and difficult 
computation) but several possibilities can be ruled out. First, the secular
dependence we found was not introduced through the gauge fixing functional. 
We found ``infrared logarithms''; these are logarithms of the ratio of the 
scale factor at the observation time to its value at the beginning of 
inflation. These logarithms break the part of the 10-parameter de Sitter 
group known as dilatations; the latter generate the transformation $x^{\mu} 
\longrightarrow k \cdot x^{\mu}$ in conformal coordinates. However, our gauge 
condition preserves this symmetry. \\

$\bullet$ {\it Infrared Logarithms Do Not Always Drop Out Of Physical And
Gauge Invariant Quantities:} The only invariant so far checked is the 
inflationary power spectrum and Weinberg has shown that infrared logarithms 
{\it do correct this} \cite{SW1,Bua}. Because the logarithms in that case 
are of the ratio of the current scale factor to the scale factor at horizon 
crossing, their enhancement is at most about $\ln(a/a_{\rm hc}) \sim 100$ 
for any perturbation whose spatial variation we can resolve. This is not 
enough to overcome the small loop counting parameter of $G H^2 \ltwid 
10^{-12}$. However, there can be arbitrarily large infrared logarithms 
correcting things we perceive as spatially constant {\it such as the vacuum 
energy}.

It might still be that the vacuum energy is somehow protected from the 
secular effects which contaminate 1PI functions \cite{TW4,MW1}, the 
quantum-corrected mode functions \cite{MW2} and the power spectrum 
\cite{SW1,Bua}. But that is certainly not true for scalar models on 
non-dynamical, de Sitter background. In that case the calculations are 
vastly simpler, and there is no gauge issue to frustrate drawing obvious 
physical conclusions. It suffices to compute the expectation value of the 
stress tensor. For example, a fully renormalized, 2-loop computation of the 
stress tensor of a massless, minimally coupled scalar with a $\varphi^4$ 
self-interaction which is released in Bunch-Davies vacuum at $t=0$ (with 
scale factor $a(t) = e^{H t}$) reveals the following secular growth for 
the induced energy density and pressure \cite{OW}:
\begin{eqnarray}
\rho(t) & = & \frac{\lambda H^4}{(2 \pi)^4} \; 
\Bigl\{ \frac18 \ln^2(a) \Bigr\} + O(\lambda^2) \; , \\
p(t) & = & \frac{\lambda H^4}{(2 \pi)^4} \; 
\Bigl\{ -\frac18 \ln^2(a) -\frac1{12} \ln(a) \Bigr\} + O(\lambda^2) \; .
\end{eqnarray}
At leading logarithm order this result is in perfect agreement with the
non-perturbative analysis of Yokoyama and Starobinski\u{\i} \cite{SY,TW7}. 
It also has the transparent physical interpretation that the inflationary
production of scalars increases the scalar field strength, which drives the
scalar up its $\varphi^4$ potential and thereby induces a growing vacuum
energy.

One may dismiss the $\varphi^4$ result on the grounds that it shows an 
{\it increase} in the vacuum energy, whereas we claim a decrease for quantum 
gravity. But it is simple to find scalar models whose induced vacuum energy 
is negative. Consider scalar quantum electrodynamics, for which the analogous 
2-loop results are \cite{RPW2,PTsW1}:
\begin{eqnarray}
\rho(t) & = & \frac{e^2 H^4}{(2 \pi)^4} \times 
\left[ -\frac34 \ln(a) \right] + O(e^4) \; , \\
p(t) & = & \frac{e^2 H^4}{(2 \pi)^4} \times 
\left[ \, \frac34 \ln(a) \right] + O(e^4) \; .
\end{eqnarray}
The physical interpretation seems to be that the inflationary production
of charged scalars polarizes the vacuum \cite{PTW,PW0}, which lowers 
the scalar energy.

Again one may dismiss the scalar QED results on the grounds that they 
represent a transient effect that approaches a constant after a 
non-perturbatively large time $\Delta t \sim 1/(e^2 H)$ \cite{PTsW1}. 
The specious nature of this argument can be seen from a massless fermion 
which is Yukawa coupled to a massless, minimally coupled scalar on 
non-dynamical de Sitter background. For that model the induced vacuum 
energy {\it falls without bound} \cite{MW3}. This also has a transparent
physical interpretation: the inflationary production of scalars increases
the scalar field strength, which induces a fermion mass, whose effect is
to decrease the vacuum energy. Had gravity been dynamical in this model, 
the resulting secular back-reaction would {\it end inflation}. That the
universe subsequently decays to a Big Rip singularity does not alter the
model's demonstration that {\it secular back-reaction can accumulate to 
dominate late time cosmology.}

Finally, it should be noted that the absence of secular quantum gravitational
back-reaction would not only require that a single infrared logarithm drops
out at two loops. The number of infrared logarithms grows with each loop in
a way that can be predicted from the number of undifferentiated massless,
minimally coupled scalars or graviton fields in the basic interaction vertex.
The number of extra infrared logarithms for each extra coupling constant in 
$\lambda \varphi^4$, Yukawa theory and scalar QED are \cite{PTsW1}:
\begin{eqnarray}
\lambda \varphi^4 \sqrt{-g} & \Longrightarrow & \ln^2(a) \qquad {\rm 
for\ every} \qquad \lambda \; , \\
-f \varphi \overline{\psi} \psi \sqrt{-g} & \Longrightarrow & \ln(a) \qquad
{\rm for\ every} \qquad f^2 \; , \\
i e \Bigl(\varphi^* \partial_{\mu} \varphi - \partial_{\mu} \varphi^*
\varphi\Bigr) A_{\nu} g^{\mu\nu} \sqrt{-g} & \Longrightarrow & \ln(a) \qquad
{\rm for\ every} \qquad e^2 \; .
\end{eqnarray}
These expectations are verified in a large number of explicit 1-loop,
2-loop, and even 3-loop computations \cite{Lots}. They follow as well from 
the stochastic analysis of Starobinski\u{\i} and Yokoyama \cite{SY,TW7,PTsW1}.
\footnote{Certain 1PI functions have counterterms that can be used to absorb 
infrared logarithms at low loop orders \cite{DKW} but factors of $\ln(a)$  
always show up at higher orders.}

The basic interaction of quantum gravity takes the form $\kappa^n h^n \partial 
h \partial h a^{D-2}$ \cite{TW9}, where $\kappa^2 \equiv 16 \pi G$ is the 
loop counting parameter. The number of extra infrared logarithms for each
extra factor of $G H^2$ is \cite{PTsW1}:
\begin{eqnarray}
\kappa^n h^n \partial h \partial h a^{D-2} & \Longrightarrow & \ln(a) \qquad
{\rm for\ every} \qquad G H^2 \; , \label{grav}
\end{eqnarray}
So the absence of secular back-reaction in quantum gravity would require 
that the escalating series of infrared logarithms {\it which do appear in 
1PI functions and physical quantities} somehow contrives to cancel out of 
the expansion rate. We do not regard it as reasonable to suppose that
this happens. Of course Garriga and Tanaka claim to have proven that it 
does happen, and to all orders. We turn now to an explanation of why their 
proof is not correct.

\section{Renormalizing the Ricci Scalar}

The basic argument of Garriga and Tanaka consists of a peculiar scheme for 
renormalizing the Ricci scalar so that the following relation applies: 
\begin{equation}
\Bigl\langle R_{\rm ren} \sqrt{-g} \, \Bigr\rangle = 
4 \Lambda \, \Bigl\langle (1 \!+\! \delta_{\rm vol}) \sqrt{-g} \, 
\Bigr\rangle \; . \label{GTrel}
\end{equation}
We shall presently describe some technical problems with this scheme but
the most significant point of this section is that (\ref{GTrel}) {\it 
does not imply a constant Ricci scalar}. We shall demonstrate that 
(\ref{GTrel}) is completely consistent with the relaxation mechanism 
of Section 1. Moreover, $R_{\rm ren}$ as defined by Garriga and Tanaka
is not a properly renormalized composite operator because it fails to
produce finite results when inserted in 1PI functions. \\

$\bullet$ {\it Preliminaries:} Many people have concluded that screening 
is impossible in pure quantum gravity because the trace of the classical 
field equations implies $R = 4 \Lambda$:
\footnote{We first heard the argument from L. Susskind in the mid 1980's 
\cite{Lenny}.}
\begin{equation}
\mathcal{L} = \frac1{16 \pi G} \Bigl(R - 2 \Lambda\Bigr) \sqrt{-g}
\;\; \Longrightarrow \;\; 
16 \pi G g^{\mu\nu} \frac{\delta S}{\delta g^{\mu\nu}}
= \Bigl(-R + 4 \Lambda \Bigr) \sqrt{-g} = 0 \; .
\end{equation}
The conclusion is premature because quantizing general relativity requires
adding to the Lagrangian an infinite sequence of ever-higher dimensional,
BPHZ counterterms:
\begin{equation}
\Delta \mathcal{L} = \alpha_1 R^2 \sqrt{-g} + \alpha_2 C^{\rho\sigma\mu\nu}
C_{\rho\sigma\mu\nu} \sqrt{-g} + \dots
\end{equation}
The divergent parts of the $\alpha_i$'s are fixed by the need to cancel
primitive divergences, but no physical principle seems to fix their finite
parts, and what they are affects physical predictions. That is one way of 
expressing the problem of quantum gravity.

Of particular interest to us is the fact that even the lowest order 
counterterms alter the relation between $R$ and $\Lambda$ implied by the 
trace of the renormalized field equations:
\footnote{Computing the trace in dimensional regularization would result 
as well in finite contributions from the $\alpha_1$ and $\alpha_2$ counterterms whose inclusion would only strengthen the argument we shall make.}
\begin{equation}
0 = 16 \pi G g^{\mu\nu} \frac{\delta (S \!+\! \Delta S)}{\delta g^{\mu\nu}(x)} 
= \Bigl(-R + 4 \Lambda + 16 \pi G \times 6 \alpha_1 \square R + \dots \Bigr)
\sqrt{-g} \; . \label{trace}
\end{equation}
One would typically break the $\alpha_i$'s up into divergent and finite
parts, and render this equation in terms of suitably renormalized composite
operators:
\begin{equation}
0 = \Bigl(-R \sqrt{-g} + 4 \Lambda \sqrt{-g} \, \Bigr)_{\rm ren} + 16 \pi G 
\times 6 \alpha_{\rm 1,finite} \Bigl( \square R \sqrt{-g}\, \Bigr)_{\rm ren} 
+ O(G^2) \; . \label{renorm}
\end{equation}
The Eddington counterterm --- the one proportional to $\alpha_1$ --- was 
exploited by Starobinski\u{\i} in constructing an early model of what would 
later be called inflation \cite{AAS1}. It should therefore be obvious that 
the equations of quantum gravity cannot imply the Ricci scalar is constant
unless the finite parts of a countably infinite number of $\alpha_i$'s are 
set to zero. \\

$\bullet$ {\it The Garriga and Tanaka Argument}: They propose to circumvent
the problem of counterterms by absorbing them into the renormalized Ricci
scalar $R_{\rm ren}$ and the renormalized volume element $(1 \!+\!
\delta_{\rm vol}) \sqrt{-g}$. Their equation (36) defines the operator whose
vanishing in expectation values is claimed to preclude secular back-reaction:
\begin{equation}
\mathcal{O}_{\rm GT} \equiv \int d^4x \, W(x) \Bigl[ -R_{\rm ren} +
4 \Lambda (1 \!+\! \delta_{\rm vol}) \Bigr] \sqrt{-g} \; . \label{dumb}
\end{equation}
Here $W(x)$ is their ``window function'' which is supposed to be a scalar
that does not depend upon the metric.
\footnote{We remark that $W(x)$ must involve the metric if it is not constant.
Garriga and Tanaka seem to feel that metric dependence in $W(x)$ would prevent
the equations of motion from holding, but any obstruction of this sort can be
absorbed into an operator ordering \cite{TW10}.}
Combining equations (38) and (41) of Garriga and Tanaka results in their
definition for $R_{\rm ren}$:
\begin{equation}
R_{\rm ren} \equiv R - 16 \pi G \times \frac{g^{\mu\nu}}{\sqrt{-g}}
\frac{\delta \Delta S}{\delta g^{\mu\nu}} + 4 \Lambda \delta_{\rm vol} \; .
\label{dumber}
\end{equation}
Substituting (\ref{dumber}) in (\ref{dumb}) reveals the key operator
of Garriga and Tanaka as the smeared trace of the renormalized field
equations:
\begin{equation}
\mathcal{O}_{\rm GT} = 16 \pi G \times \int d^4x \, W(x) \, g^{\mu\nu}(x)
\frac{\delta (S \!+\! \Delta S)}{\delta g^{\mu\nu}(x)} \; .
\end{equation}
Of course this does vanish, {\it as do all components of the renormalized
field equations}, in all expectation values, with or without the smearing. \\

$\bullet$ {\it Consistency With Screening:} Garriga and Tanaka have not 
shown their definition of the renormalized Ricci scalar is constant but 
rather that it, times the volume element, equals another operator:
\begin{equation}
\Bigl\langle R_{\rm ren} \sqrt{-g} \, \Bigr\rangle = 
4 \Lambda \, \Bigl\langle (1 \!+\! \delta_{\rm vol}) \sqrt{-g} \, 
\Bigr\rangle \; . \label{theireqn}
\end{equation}
{\it This is completely consistent with the secular screening mechanism 
described in Section 1.} It is incorrect to think of $R$ and $\sqrt{-g}$
as possessing distinct time dependences in (\ref{theireqn}) but one can 
do so at leading logarithm order. In that case, working in conformal 
coordinates, the mechanism of Section 1 implies the following relaxations
for the Ricci scalar and volume element:
\begin{eqnarray}
R = 
4\Lambda + 8\pi G ( \rho - 3p ) & \sim &
4\Lambda \Big[ 1 - \# G^2 H^4 \ln (a) + O(G^3) \Big]
\; , \label{Rrel} \\
\sqrt{-g} & \sim &
a^4 \Big[ 1 - \# G^2 H^4 \ln^2 (a) + O (G^3) \Big]
\; . \label{Volrel}
\end{eqnarray}
At leading logarithm order, the change in $R$ is insignificant compared 
with that of the volume element and both sides of equation (\ref{theireqn})
are equal:
\begin{eqnarray}
\Bigl\langle R_{\rm ren} \sqrt{-g} \, \Bigr\rangle 
& = &
4 \Lambda \, \Bigl\langle
(1 \!+\! \delta_{\rm vol}) \sqrt{-g} \, \Bigr\rangle \; , 
\nonumber \\
& \sim &
4 \Lambda \, a^4 \, \Bigl\{
1 - \# \Big[G H^2 \ln(a)\Bigr]^2 + O(G^3) \Bigr\} \; . 
\label{oureqn}
\end{eqnarray}
This continues to be true at all orders in the loop expansion, as per
relation (\ref{grav}), because $R$ involves differentiated metrics
whereas $\sqrt{-g}$ does not. Of course Garriga and Tanaka want equation 
(\ref{theireqn}) to be exact and not just valid at leading logarithm 
order. However, the necessary sub-leading logarithms can easily be 
supplied by cross terms between $R$ and $\sqrt{-g}$, and by the operator 
$\delta_{\rm vol}$. \\

$\bullet$ {\it Peculiar Renormalization:} Although the renormalization scheme 
(\ref{dumber}) is consistent with secular slowing, we still find it dubious. 
As Garriga and Tanaka point out, using the equations of motion inside a 
functional integral requires the integrand to be an invariant \cite{TW10}. 
Hence one is not considering ``$R$'' and ``$1$'' but rather ``$R \sqrt{-g}
\, $'' and ``$\sqrt{-g} \, $''. Renormalizing such composite operators requires that their insertion in a 1PI diagram be accompanied by the insertion of 
counter-operators:
\begin{eqnarray}
\Bigl( R \sqrt{-g} \, \Bigr)_{\rm ren} & \equiv & R \sqrt{-g} + \delta R
\sqrt{-g} \; , \\
\Bigl( \sqrt{-g} \, \Bigr)_{\rm ren} & \equiv & \sqrt{-g} + \delta_{\rm vol}
\sqrt{-g} \; .
\end{eqnarray}
We know of no previous study of these operators with a non-zero cosmological
constant but, on general grounds of symmetry and the degree of 1-loop 
divergences one expects \cite{SW2,IZ}:
\begin{eqnarray}
\delta_{\rm vol} & \!\!=\!\! & 
16 \pi G \times \gamma_1 R 
+ O(G^2) \; , \label{dV} \\
\delta R & \!\!=\!\! & 
16 \pi G \times \Bigl(
\beta_1 \square R + \beta_2 R^2 + \beta_3 R^{\mu\nu} R_{\mu\nu} 
+ \beta_4 C^{\rho\sigma\mu\nu} C_{\rho\sigma\mu\nu} 
+  4 \gamma_1 \Lambda R \Bigr) 
\qquad \nonumber \\
& \mbox{} & 
+ O(G^2) \; . 
\end{eqnarray}
The trace of the renormalized field equations (\ref{trace}) requires that
the divergent part of $\beta_1$ should agree with the divergent part of
$-6 \alpha_1$, and that $\beta_{2-4}$ should vanish. That is no problem.
However, the renormalization scheme (\ref{dumber}) of Garriga and Tanaka 
also requires that the finite part of $\beta_1$ should agree with the finite 
part of $-6 \alpha_1$. An escalating series of similar relations must hold 
as one moves up the loop expansion.

There are two ways of viewing this. Either Garriga and Tanaka have solved the
problem of quantum gravity by uniquely specifying the finite parts of every
$\alpha_i$ except $\alpha_2$,\footnote{The single parameter $\alpha_2$ would 
not prevent us from making sense of quantum gravity.} or else they are 
defining the Ricci scalar so as to absorb known classical effects. In the 
latter case one could also absorb the crucial $R^2$ term of Starobinski\u{\i}'s
model \cite{AAS1} to conclude that the ``renormalized Ricci scalar'' is 
constant. Indeed, there seems no reason to restrict the procedure to pure 
gravity. Were it applied to scalar-driven inflation --- by including the 
scalar action in the ``$\Delta S$'' they use in (\ref{dumber}) --- then one 
would conclude that scalar-driven inflation never ends.

It might be objected that a key distinction between the counterterms of 
quantum gravity and the inflaton potential is that the former contain factors 
of $\hbar$ whereas the latter does not. If so, then suppose the inflaton
potential derives from a one loop correction, such as the Coleman-Weinberg
potential of New Inflation \cite{AS}. The inflaton potential now carries a
factor of $\hbar$, so it can be absorbed into the ``renormalized Ricci scalar''
and we have just shown that New Inflation is pure de Sitter. \\

$\bullet$ {\it The Composite Operator $R_{\rm ren}$ Of (\ref{dumber}) Is 
Neither Properly Defined Nor Is It Constant:} Correctly defined composite 
operators have the property that inserting them in 1PI functions gives 
finite results. With the conventions of Garriga and Tanaka, the composite 
operators $R_{\rm ren} \sqrt{-g}$ and $(1 \!+\! \delta_{\rm ren}) \sqrt{-g}$ 
have this property but $R_{\rm ren}$ does not. Simply divide out the volume 
element from both sides of their key relation (\ref{GTrel}):
\begin{equation}
R_{\rm ren} \sqrt{-g} = 4 \Lambda (1 + \delta_{\rm vol}) \sqrt{-g} 
\qquad \Longrightarrow \qquad 
R_{\rm ren} = 4 \Lambda (1 + \delta_{\rm vol}) \; .
\end{equation}
Now note two crucial features of the volume counterterm $\delta_{\rm vol}$:
\begin{enumerate}
\item{It isn't finite; and}
\item{It is an operator rather than a constant.}
\end{enumerate}
The first feature can be seen by expanding the volume element operator
in powers of the graviton field:
\begin{equation}
\sqrt{-g} = a^D \Bigl(1 + \frac{\kappa}2 h^{\rho}_{~\rho} + 
\frac{\kappa^2}8 (h^{\rho}_{~\rho})^2 - \frac{\kappa^2}4 h^{\rho\sigma} 
h_{\rho\sigma} + \dots \Bigr) \; .
\end{equation}
Inserting $\sqrt{-g}$ in the vacuum amplitude gives a divergence at
order $\kappa^2 = 16 \pi G$ from the coincident propagator.
\footnote{The divergence would be $\frac{\kappa^2 H^2}{4 \pi^2} \frac1{D-4}$ 
in our gauge, implying $\gamma_1 = -\frac1{48 \pi^2} \frac1{D-4} + 
{\rm finite}$.} 
From expression (\ref{dV}) we see that this divergence is canceled by 
the 1-loop contribution to $\delta_{\rm vol}$, which is
$\kappa^2 \gamma_1 R = \gamma_1 D (D\!-\!1) \kappa^2 H^2 + O(\kappa^4)$.
It follows that $\gamma_1$ is divergent, so inserting just $\delta_{\rm vol}$ 
into a 1PI function, {\it without the factor of} $\sqrt{-g}$, does not 
produce finite results. We therefore conclude that Garriga and Tanaka cannot 
claim to have shown that the Ricci scalar is constant --- or anything else 
--- because they have not correctly defined it as a composite operator. 
Further, the quantity $R_{\rm ren}$, is not constant but rather an infinite 
series of time-dependent operators. 

\section{Ricci Scalar Constancy and Expansion}

It is time to critically examine the unstated assumption of Garriga and
Tanaka that proving the Ricci scalar is constant implies de Sitter expansion. 
This is quite incorrect. Neither the classical field equation, $R_{\mu\nu} - 
\frac12 R g_{\mu\nu} + \Lambda g_{\mu\nu} = 0$, nor its trace, $-R + 4 
\Lambda = 0$, precludes the presence of gravitational radiation. In fact, 
all gravitational wave solutions with $\Lambda \neq 0$ must obey these 
equations. Because gravity is a nonlinear theory, gravitational radiation
interacts with itself. Because the gravitational force is attractive,
this self-interaction opposes the expansion of spacetime. If enough 
gravitational radiation is present, the result is not expansion but rather 
collapse into one or more black holes. This is an unavoidable consequence 
of the singularity theorems. Therefore, one cannot conclude that constant
Ricci scalar implies a constant expansion rate because this is not even
true classically.

It might be objected that a positive cosmological constant opposes the
tendency towards collapse of too much gravitational radiation. So the
correct result is that a fixed, initial distribution of gravitational 
radiation may well collapse to one or more black holes, but the late time 
geometry away from these black holes will approach the local de Sitter 
form for which $R^{\rho}_{~\sigma \mu\nu} = H^2 (\delta^{\rho}_{\mu} 
g_{\sigma \nu} - \delta^{\rho}_{\nu} g_{\sigma \mu})$. That would be 
true enough classically, where no new gravitational radiation can be
generated beyond what was present initially, but it is not correct when
quantum effects are included. The simplest way to understand quantum
effects is as the classical response to the source provided by uncertainty
principle. In these terms, our screening mechanism is driven by the {\it 
steady injection} of gravitational radiation, throughout space, as more 
and more infrared virtual gravitons are ripped out of the vacuum. These
gravitons must slow the expansion of spacetime because that is what a 
classical distribution of gravitational radiation would do. However, 
such a classical distribution would obey $R = 4 \Lambda$.

\section{Epilogue}

It seems obvious that gravitons contribute to vacuum energy. At 1-loop 
order, where each mode contributes separately, infrared gravitons participate 
on an equal footing with ultraviolet ones; this can be shown by explicit
computation \cite{Ford,Fabio,TW3}. Higher loop contributions involve non-linear 
combinations of gravitons which are best viewed as integrals over position 
space. These integrals reach back from the point of observation, along the 
past light-cone to the fixed initial value surface. Because the past light-cone grows as the time of observation evolves, these higher loop contributions can 
grow as well. That has to be expected in view of the infrared singularities of
quantum gravity on a locally de Sitter background \cite{TW11}.

Nothing ought to seem dubious about this. Although the effect has not 
yet been demonstrated for an invariant measure of the quantum gravitational 
back-reaction on inflation, the same phenomenon injects secular dependence
into other invariant quantities \cite{SW1,Bua}. It also engenders secular 
contributions to the vacuum energies in scalar models \cite{OW,PTsW1,MW3}
for which there is no gauge issue to prevent one from drawing obvious
conclusions.

We have shown that the renormalization scheme employed by Garriga and Tanaka
is highly dubious and that, even if accepted, it does not preclude secular
slowing. Nor would proving any plausibly defined ``Ricci scalar'' to be
constant preclude secular slowing. Of course the burden of proving that 
secular slowing does occur rests quite properly with us. And even an explicit 
perturbative demonstration of this at 2-loop order would not establish 
quantum gravitational back-reaction either as an explanation for why the 
observed cosmological constant is so small or as the basis of a viable 
model of inflation. For that one would need a reliable way of computing in 
the late time regime, after back-reaction has become non-perturbatively strong.

We do not think such computations are beyond reach. Starobinski\u{\i} has
proposed a simple stochastic formalism \cite{AAS2} that reproduces the
leading infrared logarithms of scalar potential models to all orders
\cite{TW7} and which can be summed to give non-perturbative results
\cite{SY}. We have recently extended Starobinski\u{\i}'s formalism to
massless, minimally coupled scalars that interact with other kinds of
fields. This has led to explicit, non-perturbative results for Yukawa
theory \cite{MW3} and for scalar QED \cite{PTsW1}. We do not yet know
how to treat derivative interactions of the sort quantum gravity possesses
but the problem does not seem insolvable.

\vskip .7cm

\centerline{\bf Acknowledgements}

We thank J. Garriga and T. Tanaka for friendly correspondence on these
fascinating issues. This work was partially supported by the European
Social fund and National resources $\Upsilon\Pi{\rm E}\Pi\Theta$-Pythagoras
II-2103, by European Union grant FP-6-12679, by NSF grant PHY-0653085, and
by the Institute for Fundamental Theory at the University of Florida.

\end{document}